\documentclass[acmtog]{acmart}
\usepackage{algorithm}
\usepackage{algpseudocode}

\usepackage{amsmath,amssymb}

\setcopyright{cc}
\setcctype{by}
\copyrightyear{2026}
\acmYear{2026}
\acmDOI{10.1145/3811351}

\acmJournal{TOG}
\acmVolume{45}
\acmNumber{4}
\acmArticle{39}
\acmMonth{7}

\acmSubmissionID{877}

\newcommand{\revise}[1]{#1}
\newcommand{\method}[1]{{\textcolor{black}{\emph{HoloPathTracer}#1}}}

\usepackage{xcolor}
\usepackage{xspace}
\usepackage{colortbl}

\definecolor{colorDomainA}{rgb}{0.95, 0.95, 0.75}
\definecolor{colorDomainB}{rgb}{0.95, 0.95, 0.75}

\definecolor{colorTrd}{rgb}{1, 0.8, 0.8}
\definecolor{colorSnd}{rgb}{1, 0.90, 0.80}
\definecolor{colorFst}{rgb}{0.85, 0.95, 0.85}

\usepackage{enumitem}
\usepackage{siunitx}
\usepackage{pifont}
\begin{document}

\title{HoloPathTracer: Fast and Accurate Wave Path Tracing for Holography}

\author{Wenbin Zhou}
\authornote{denotes equal contribution.}
\orcid{0009-0001-4487-2877}
\affiliation{%
  \institution{The University of Hong Kong}
%  \city{Hong Kong}
  \country{Hong Kong SAR}}
\email{zhouwb@connect.hku.hk}

\author{Xiangyu Meng}
\authornotemark[1]
\orcid{0009-0002-8828-7036}
\affiliation{%
  \institution{The University of Hong Kong}
%  \city{Hong Kong}
  \country{Hong Kong SAR}}
\email{mengxy22@connect.hku.hk}

\author{Jiankai Xing}
\orcid{0000-0002-8341-9952}
\affiliation{%
  \institution{The University of Hong Kong}
%  \city{Hong Kong}
  \country{Hong Kong SAR}}
\affiliation{%
  \institution{Tsinghua University}
%  \city{Beijing}
  \country{China}}
\email{xjk21@mails.tsinghua.edu.cn}

\author{Xin Liu}
\orcid{0000-0002-4371-018X}
\affiliation{%
  \institution{The University of Hong Kong}
%  \city{Hong Kong}
  \country{Hong Kong SAR}}
\email{liuxin24@hku.hk}

\author{Suyeon Choi}
\orcid{0000-0001-9030-0960}
\affiliation{%
  \institution{Stanford University}
%  \city{Palo Alto}
  \country{USA}}
\affiliation{%
  \institution{Seoul National University}
%  \city{Seoul}
  \country{Republic of Korea}}
\email{suyeon@stanford.edu}

\author{Yifan Peng}
% \authornote{Corresponding author.}
\orcid{0000-0003-0667-2599}
\affiliation{%
  \institution{The University of Hong Kong}
%  \city{Hong Kong}
  \country{Hong Kong SAR}}
\email{evanpeng@hku.hk}

% \renewcommand{\shortauthors}{Wenbin Zhou, Xiangyu Meng, Jiankai Xing, Xin Liu, Suyeon Choi, and Yifan Peng}

%%

\begin{abstract}
Holography offers unique advantages for delivering perceptual realism while preserving compact form factors in VR/AR. Its perceptual quality, however, hinges on encoding rich wavefronts of photorealistic scenes into interference patterns and then incoherently multiplexing the resulting wave fields for perception.
Existing CGH paradigms decouple radiance estimation from wave propagation by pre-rendering radiance on discretized scene sectors. This separation between radiometric and wave-optical computation inherently limits the range of focus cues and visual effects that can be faithfully reproduced, including depth- and view-continuity, and physically based material behaviors such as glossy or mirror-like reflection and refraction.

We present a \emph{physically accurate yet computationally efficient} wave optics rendering framework leveraging path tracing to encode full 3D visual cues into phase holograms.
Specifically, we employ a Monte Carlo method to solve both the rendering equation and the Rayleigh--Sommerfeld integral simultaneously.
Our algorithm is fully compatible with modern graphics techniques and can generate multiple time-multiplexed random holograms with minimal additional time cost via \textit{Path Reuse}. By employing a fast approximation with an ambient radiance cache, we realize an order of magnitude convergence speed improvement.
The resulting coherent wave fields that inherently encode comprehensive visual effects are converted into phase-only holograms under complex-amplitude supervision.
Through extensive simulations and experimental validations on a spatial light modulator-based display prototype, we demonstrate faithful holographic reconstructions of natural 3D cues and complex materials, including realistic defocus blur, view-dependent effects, as well as appearance highlights and reflections.
  \vspace{-2pt}
\end{abstract}



\begin{teaserfigure}
  \includegraphics[width=\textwidth]{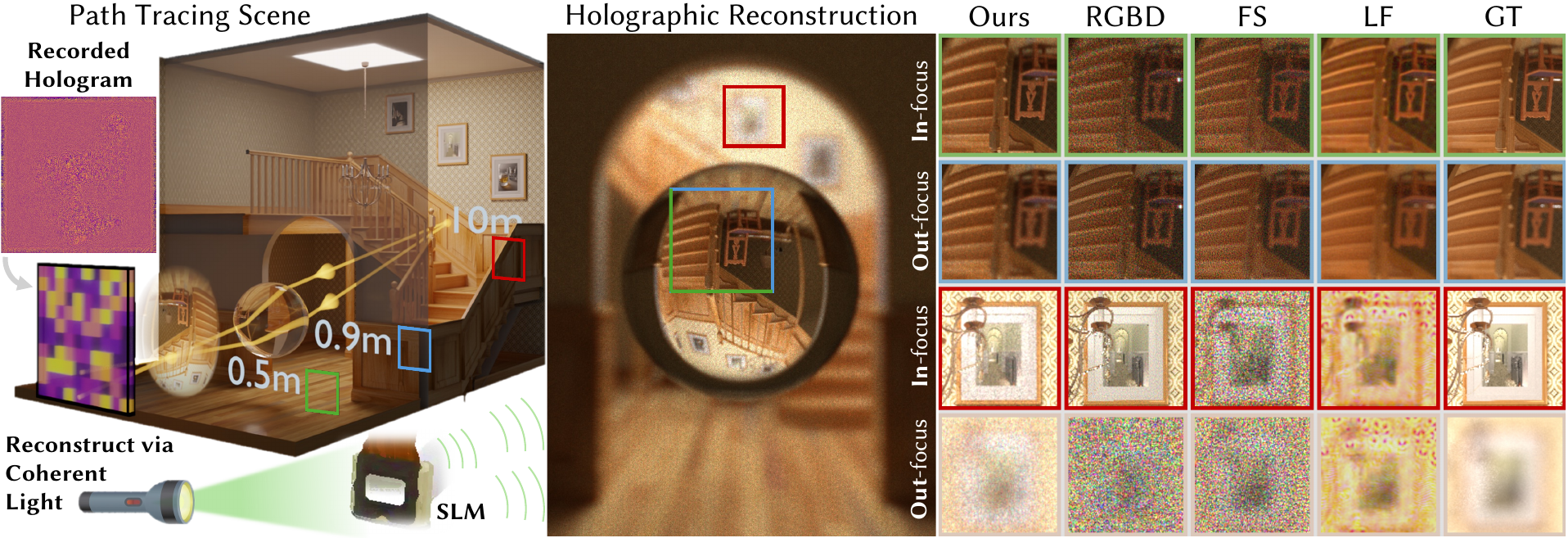}
  \vspace{-20pt}
  \caption{Technical footprint and reconstruction results of our HoloPathTracer. \textbf{Left}: Wave field rendering from a target scene to the hologram recording plane, with physically accurate visual cues encoded. The resulting hologram is displayed on an SLM and illuminated by coherent light for optical reconstruction.
  \textbf{Center}: Reconstruction (simulation) via propagating the hologram-modulated wavefront to the observation plane, with viewpoint near-focused at the imagery of the stairs via the glass sphere.
  \textbf{Right}: Close-up comparisons of our method and existing CGH baselines, validating superior visual fidelity (e.g., natural defocus, glossy highlight) at varying physically accurate depths (0.5m, 0.9m, and 10m). FS: Focal stack; LF: Light field; GT: Ground truth from Mitsuba3.
  }
  \Description{Composite teaser showing the HoloPathTracer workflow from wave field rendering to SLM display and optical reconstruction, followed by simulated reconstructions and zoomed comparisons against CGH baselines at multiple depths.}
  \label{fig:teaser}
\end{teaserfigure}  


\maketitle

%%%%%%%%%%%%%%%%%%%%%%%%%%%%

\section{Introduction}
\enlargethispage{.30\baselineskip}
The objective of holographic displays is to reconstruct a wavefront indistinguishable from those emitted by real objects, thereby enhancing the immersive capabilities of VR/AR~\cite{jang2024waveguide, zhou2025empowering}.
By leveraging light diffraction and interference, such displays typically employ a spatial light modulator (SLM) with a synthesized 3D hologram to manipulate coherent wave fields~\cite{blanche2021holography}.
Realizing \emph{Perceptual Realism}---with natural depth and other visual cues---motivates mimicking incoherent emission via multiplexing coherent behaviors inherent in holography (see also Sec.~3.3), a direction further enabled by emerging high-speed SLMs~\cite{choi2022time}.
Complementing photonics hardware advances~\cite{yin2022advanced}, computer-generated holography (CGH) algorithms compute the wave fields on the hologram plane from a 3D scene, which are then encoded into phase-only holograms through either iterative or direct methods~\cite{choi2021neural} for optical reconstruction.

Exploring a \textbf{physically accurate yet computationally efficient rendering framework for complex-valued wave fields} is essential for realistic handling of specular reflection, glossy surfaces, and transparent materials under in-focus and defocused conditions.
Yet, this rendering, grounded in Rayleigh--Sommerfeld (RS) and Huygens principles~\cite{shen2006fast,buitrago2019non}, is computationally demanding due to stringent sampling requirements and the heavy load of multi-view propagation.
To remain tractable, existing numerical diffraction computations for CGH ``practically'' abandon fully physically accurate wave propagation models in favor of \emph{approximate} formulations.

Wave propagation formulated with point-based models~\cite{tsang2018review}, while physically expressive, demands dense sampling and becomes prohibitively expensive when rendering coherent fields for complex scattering and involving multiple viewing angles.
Alternative image/layer-based models (e.g., focal stacks, holographic stereograms, light fields~\cite{choi2022time,schiffers2023stochastic}) partition the scene into discretized layers and leverage FFT
to accelerate intra-layer propagation.
Yet, the inter-layer \emph{discretization} compromises perceptual quality (e.g., incorrect scene occlusions), and runtime still scales linearly with the number of layers.
Polygon-based paradigms~\cite{wang23high} similarly adopt layer-based propagation, 
\revise{where high-fidelity configurations are achieved, yet computation scales linearly with the facet count.}
Recently, Gaussian primitives-based propagation enables fast neural rendering of 3D cues~\cite{choi2025gaussian}. This concurrent direction remains an \emph{approximation} to full physical wave propagation and generally yields smooth-like patterns, trading reconstruction fidelity (e.g., correct refraction/reflection) for computational efficiency, akin to prior formulations.
More broadly, overly smooth patterns may suppress the stochastic nature of holograms, which is important for resilience and perceptual realism, as discussed in Sec.~\ref{sec:smoothrandomtradeoff}.

Path tracing is a physically accurate framework for modeling light transport, as it characterizes the propagation of light energy in a scene by stochastically sampling light paths via sequences of surface interactions~\cite{yu2023full,steinberg2024generalized}.
Integrating such a rendering paradigm with wave optics in an efficient manner is of significance for advancing CGH~\cite{watanabe2024realistic}.
Notably, conventional path tracing pipelines render only radiance (amplitude) while discarding the phase information carried by light.
Consequently, they do not ``encode'' the full complex wavefront essential for holographic reconstruction.
Yet, naively computing the complex wavefront via explicit wave propagation remains computationally prohibitive~\cite{chen2023photorealistic,blinder2021photorealistic}, due to multi-bounce light transport and high sampling rates demanded.

This work employs RS integral-grounded path tracing to render complex-valued wave propagation, without relying on predefined, discretized geometric primitives.
The resulting wave field naturally encodes physically based rendering effects such as view-dependent highlights, specular reflections, transmissive materials, and global illumination, enabling holographic reconstructions with photorealism comparable to that of the Mitsuba renderer~\cite{jakob2022mitsuba3}.
To validate the rendered wave field, we adopt a stochastic gradient descent-based phase-encoding algorithm supervised in the complex-valued domain, and build a holographic display prototype using a phase-only SLM-based light engine.

In summary, this work makes the following contributions:
\begin{itemize}[label=\raisebox{0.05ex}{\scalebox{1.55}{$\bullet$}},labelindent=1em,leftmargin=*,labelsep=0.55em,itemsep=0.25em,topsep=0.2em]
    \item We propose an explicit wave path tracing rendering framework for CGH, \method, that computes physically accurate complex-valued wave fields for 3D scenes. Our pipeline decomposes physically based illumination into coherent and incoherent components, pre-optimizing the incoherent part, to preserve physical fidelity while realizing practical computational efficiency.

    \item We develop a parallel-like rendering pipeline that efficiently generates random-like instances of the complex wave field to support multi-frame time-multiplexed holographic displays.

    \item Our method exhibits sub-linear scaling with respect to the number of primitives in 3D scenes. Consequently, we successfully conduct simulations on highly complex physics-based rendering mesh scenes with up to \emph{two million} geometric elements, maintaining photorealism comparable to Mitsuba path-traced renderings.

    \item We build a holographic display prototype, optimize phase-only holograms, and experimentally show holographic images with natural defocus cues.
\end{itemize}
\vspace{1pt}
\noindent Source code is available on our GitHub repository\footnote{\href{https://github.com/zhou-wb/HoloPathTracer}{github.com/zhou-wb/HoloPathTracer}}.

\section{Related Work}
\label{sec:relatedwork}
This work builds on recent progress in CGH algorithms, but focuses on exploring a physically accurate yet computationally efficient wave field rendering framework for CGH.
For a more comprehensive overview of holographic displays, we refer the readers to~\cite{blanche2021holography, chang2020nextgeneration, javidi2021roadmap}.

\begin{table*}[t]
\caption{\textbf{Summary of CGH algorithms w.r.t. computational efficiency and reconstruction visual fidelity}.
Compared frameworks include the point-based, polygon-based, CNN w/ RGB-D input, SGD-based w/ focal stacks and light fields, and Gaussian-based wave splatting methods.
\revise{Herein, \ding{115}/\ding{52}/\ding{55} denote partial/yes/no for each item, and green/yellow/red indicate good/moderate/poor capability.}
Our approach uses readily available 3D scene data in graphics with less dependency on pre-defined geometric primitives, and does not require pre-training any networks tied to a specific configuration.
Importantly, the path tracing-empowered wave optics rendering naturally supports advanced shading, occlusion, and view-dependent effects.
Representative methods: 1~\cite{maimone2017holographic}, 2~\cite{wang23high}, 3~\cite{shi2021realtime}, 4~\cite{choi2022time}, 5~\cite{kim2024holographic}, 6~\cite{choi2025gaussian}.
}
\vspace{-8pt}
\footnotesize
\label{tab:main_compare}
\centering
\renewcommand{\arraystretch}{1.3}
\resizebox{\textwidth}{!}
{%
\begin{tabular}{l||c|c|c|c|c|c|c|c|c} 
\hline
             & \multicolumn{4}{c|}{\textbf{Computational Efficiency}}  & \multicolumn{5}{c}{\textbf{Visual Fidelity}}   \\ 
\hline
             & Fast inference & Efficient multi-frame  & Data availability & Pretraining required  & Reflection/Refraction & Complex material/GI & View-dependent  & Depth-continuity   & Natural defocus            \\ 
\hline

Point-based \footnotemark[1]           & \cellcolor{colorSnd}\ding{115}  
& \cellcolor{colorFst}\ding{52}        & \cellcolor{colorFst}\ding{52}           
& \cellcolor{colorFst}\ding{55}        & \cellcolor{colorTrd}\ding{55} 
& \cellcolor{colorSnd}\ding{115}     
& \cellcolor{colorFst}\ding{52}        & \cellcolor{colorFst}\ding{52}  
& \cellcolor{colorTrd}\ding{55} \\
Polygon-based \footnotemark[2]         & \cellcolor{colorTrd}\ding{55}  
& \cellcolor{colorTrd}\ding{55}        & \cellcolor{colorFst}\ding{52}           
& \cellcolor{colorFst}\ding{55}        & \cellcolor{colorSnd}\ding{115}
& \cellcolor{colorSnd}\ding{115}       
& \cellcolor{colorFst}\ding{52}        & \cellcolor{colorFst}\ding{52}  
& \cellcolor{colorFst}\ding{52} \\
CNN w/ RGBD \footnotemark[3]           & \cellcolor{colorFst}\ding{52}   
& \cellcolor{colorSnd}\ding{115} & \cellcolor{colorFst}\ding{52}           
& \cellcolor{colorTrd}\ding{52}        & \cellcolor{colorTrd}\ding{55} 
& \cellcolor{colorSnd}\ding{115}         
& \cellcolor{colorTrd}\ding{55}        & \cellcolor{colorTrd}\ding{55}  
& \cellcolor{colorTrd}\ding{55}  \\
SGD w/ Focal Stack \footnotemark[4]    & \cellcolor{colorTrd}\ding{55}  
& \cellcolor{colorSnd}\ding{115} & \cellcolor{colorTrd}\ding{55}           
& \cellcolor{colorFst}\ding{55}        & \cellcolor{colorSnd}\ding{115}
& \cellcolor{colorSnd}\ding{115}       
& \cellcolor{colorTrd}\ding{55}        & \cellcolor{colorSnd}\ding{115} 
& \cellcolor{colorFst}\ding{52} \\
SGD w/ Light Field \footnotemark[5]    & \cellcolor{colorTrd}\ding{55}   
& \cellcolor{colorSnd}\ding{115} & \cellcolor{colorTrd}\ding{55}            
& \cellcolor{colorFst}\ding{55}        & \cellcolor{colorSnd}\ding{115}   
& \cellcolor{colorSnd}\ding{115}           
& \cellcolor{colorFst}\ding{52}        & \cellcolor{colorSnd}\ding{115}       
& \cellcolor{colorFst}\ding{52} \\
Gaussians-based \footnotemark[6]       & \cellcolor{colorSnd}\ding{115}  
& \cellcolor{colorTrd}\ding{55}        & \cellcolor{colorFst}\ding{52}           
& \cellcolor{colorTrd}\ding{52}        & \cellcolor{colorTrd}\ding{55}  
& \cellcolor{colorSnd}\ding{115}          
& \cellcolor{colorFst}\ding{52}        & \cellcolor{colorFst}\ding{52}   
& \cellcolor{colorTrd}\ding{55}   \\
\hline
\textbf{Path Tracing} (ours)           & \cellcolor{colorSnd}\ding{115} 
& \cellcolor{colorFst}\ding{52}        & \cellcolor{colorFst}\ding{52}           
& \cellcolor{colorFst}\ding{55}        & \cellcolor{colorFst}\ding{52} 
& \cellcolor{colorFst}\ding{52}            
& \cellcolor{colorFst}\ding{52}        & \cellcolor{colorFst}\ding{52}  
& \cellcolor{colorFst}\ding{52} \\
\hline
\end{tabular}
}
\vspace{-4pt}
\end{table*}

\paragraph{Wave Field Rendering and Hologram Recording}
Prior CGH pipe\-lines first render scene radiance on \emph{discrete} geometric primitives, and then apply numerical wave propagation to the hologram plane.
Point clouds are widely employed due to their implementation simplicity~\cite{tsang2018review,chakravarthula2019wirtinger}, but they typically require extensive sampling. Polygon meshes~\cite{wang23high,matsushima2009extremely} 
can be processed analytically, offering high computational speed at the cost of lacking detailed texture representation; or numerically, they resemble point-based methods and demand significant sampling during baking process.
Layer (RGBD)-based sampling methods are computationally attractive as they operate on a finite set of pre-rendered 2D images~\cite{choi2021neural,kavakli2023realistic}, but they can struggle to provide continuous depth cues. 
Light fields can be transformed via short-time Fourier transform (STFT) into holographic stereograms~\cite{padmanaban2019holographic,pupilaware}, which encode both depth- and view-dependent effects but remain computationally expensive.
Recently, Choi et al.~\shortcite{choi2025gaussian} utilize 2D Gaussians for CGH with enhanced computational efficiency over classical meshes-based methods.
Despite progress, initiating wave propagation from discretized, pre-rendered primitives inevitably introduces approximations that degrade wave-field accuracy and hinder faithful reproduction of certain optical/shading effects (see Table~1). This work integrates these two stages---radiance rendering and hologram wave recording---to directly derive target wave fields. 

\paragraph{Ray Tracing with Wave Optics}
Ray tracing has been widely adopt\-ed in optical system analysis and lens design~\cite{wang2022differentiable,steinberg2021generic}. 
\revise{Recently, wave-optical behaviors have been incorporated into Monte-Carlo rendering frameworks~\cite{liu2025fully,steinberg2022towards,magallon2021slm} to model effects beyond the reach of geometric optics alone, including diffraction and high-frequency interference patterns~\cite{steinberg2022rendering,kim2025monte,bar2019monte}, as well as wave-informed BRDFs~\cite{wei2025learned,zeng2025designing}. }
As a representative, Steinberg et al.~\shortcite{steinberg2024generalized} introduce generalized rays for backward wave-optical light transport, enabling sensor-to-source path sampling while retaining the coherence information needed for rendering diffraction and interference effects.

Particularly relevant to our work, Blinder et al.~\shortcite{blinder2021photorealistic} introduce a point-based CGH framework rendered via path tracing; however, their method primarily targets global illumination on point primitives and does not model coherent propagation through complex materials. 
Chen et al.~\shortcite{chen2023photorealistic} improve BSDFs and ray samplings to accelerate computation. Notably, as these methods leverage path tracing solely to acquire physically accurate radiance without carefully incorporating ``long-path'' wave propagation influence, they offer limited support for encoding full visual cues via transparent or glossy objects and demonstrate results on a narrow range of reconstruction scenes.
We primarily validate the proposed framework using the Disney principled BSDF~\cite{burley2015extending}; nevertheless, established path integral regimes that exploit wave-optical light transport and microstructure-based BxDFs~\cite{yan2018rendering,jakob2014comprehensive,steinberg2024free} are, in principle, compatible with our formulation and can be incorporated.

\paragraph{Smooth-Random Phase Trade-off and Photorealism Matter}
\label{sec:smoothrandomtradeoff}
The spatio-angular behavior of resulting reconstructions can depend on the wavefront's phase profile: smooth-phase versus random-phase holograms~\cite{yoo2021optimization,choi2025phase}.
The former type~\cite{shimobaba2015review} typically yields reasonable 2D image-like quality for a few fixed viewpoints/depths; however, the relative energy concentration in low angular frequencies inherently limits the eye-box size and produces unnatural defocus cues~\cite{tseng2024neural,xia2025multi}.
By contrast, constructing a random-like phase for a given 3D scene better utilizes the available spatio-angular bandwidth of holographic displays~\cite{schiffers2023stochastic}, a prerequisite for perceptual realism. 
This benefit comes at a cost: rapid phase variations across pixels tend to induce speckle noise and other visual artifacts~\cite{goodman2007speckle}, which can become more prominent when enforcing pixelated SLMs affected by crosstalk~\cite{ban2026towards}. Consequently, additional speckle reduction and hardware calibration strategies are typically required~\cite{peng2021specklefree}.

Faithfully reproducing visual effects requires considering both amplitude and phase during hologram recording. Yet, established solutions, including aforementioned layers- and light field-based ones, primarily enforce constraints on amplitude to approximate wavefront continuity in 3D space~\cite{shi2021realtime}, thereby only \emph{indirectly} shaping phase during modeling and ultimately achieving ``partial'' photorealism. 
Prior work that integrates path tracing concepts into CGH~\cite{blinder2021photorealistic,sun2020acceleration} has targeted either computational acceleration or vanilla global illumination, that are, again, not inherently phase-centric.
This work departs from existing wave presentations: we develop a microfacet-extended formulation and a full wave field rendering framework from light sources to hologram recording planes (see Sec.~\ref{subsection:wavefacet}).
This design provides fine-grained control over the hologram ``spectrum'' in both amplitude and phase, further enabling \emph{multiple} time-multiplexing frames to be rendered in parallel with minimal extra cost via path reuse, thereby supporting a richer set of 3D visual effects at reduced computational expense.

\paragraph{Computational Techniques for Hologram Encoding}
Direct methods (e.g., DPAC and its variants~\cite{shi2022endtoend}) are highly efficient, analytically decomposing a complex-amplitude into a coherent sum of two phase-only patterns, but at the cost of reduced light efficiency and additional opto-mechanical filtering.
Deep learning offers an alternative direct-inference route via inverse networks~\cite{liu2025propagation,zhou20253d}, some with support for compression through compact latent representations~\cite{ban2026multi}. Yet, many models tend to generate phase patterns with checkerboard-like structures, an artifact that may degrade holographic reconstructions.
Most hologram encoding schemes tend to overfit the amplitude of supervised pixels~\cite{choi2021neural}, rather than faithfully reconstructing the underlying 3D wave field. 
Our framework is compatible with mainstream phase-only hologram encoding paradigms and we adopt SGD-based optimization under complex-amplitude supervision~\cite{Chen2021multidepth}. By explicitly constraining both amplitude and phase of the target field, we mitigate the tendency of overfitting to pixel-wise intensity. We also introduce a phase-shift-invariant scale factor to handle global and local phase ambiguities for better convergence.

\section{Preliminaries of Path Tracing and Wave Propagation}

\begin{figure*}[htb]
  \includegraphics[width=1\textwidth]{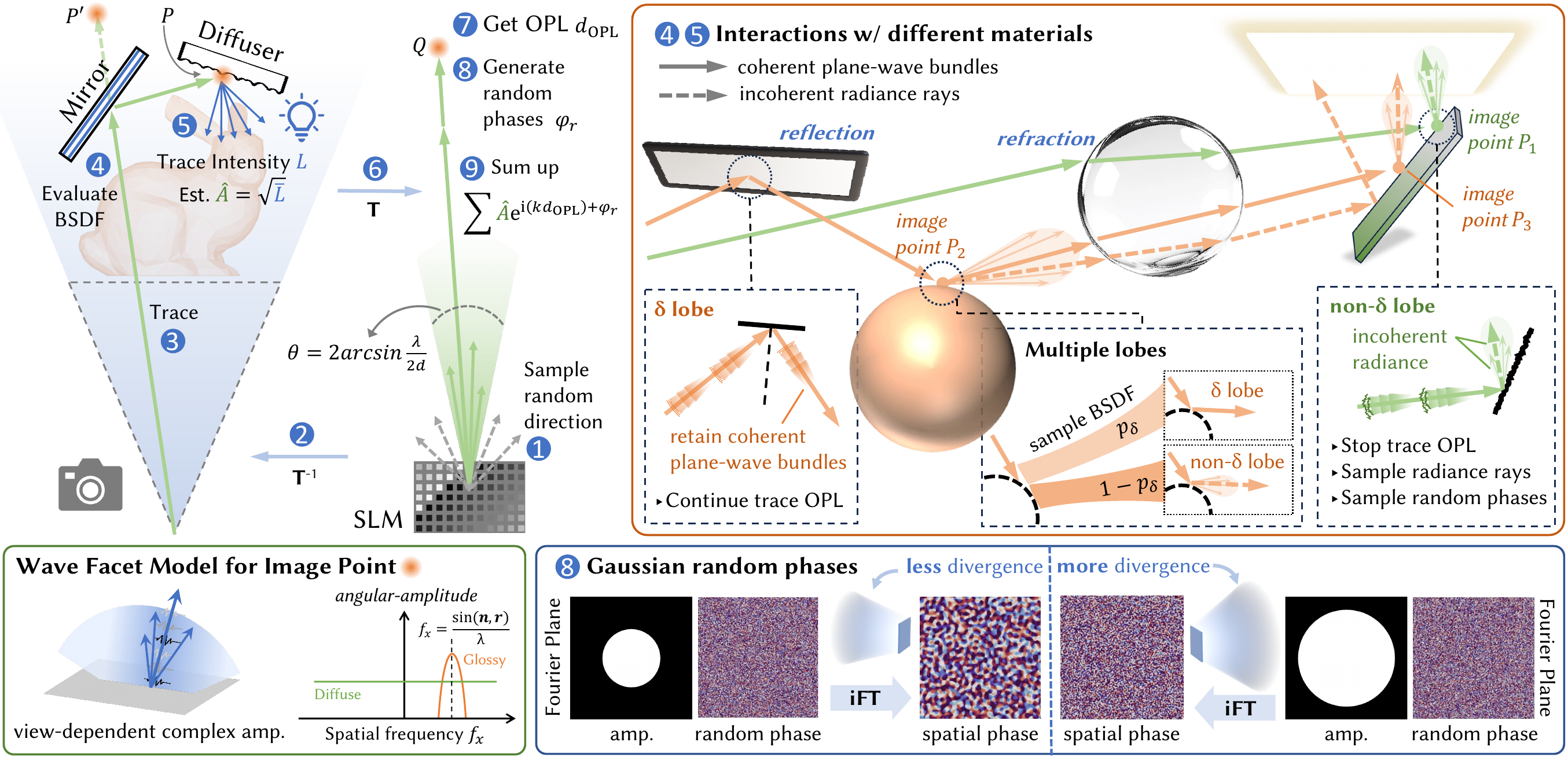}
  \vspace{-15pt}
  \caption{\textbf{\revise{HoloPathTracer processing pipeline}}. \revise{\textbf{Top-left:} the proposed method can be divided into \emph{9 Stages} including ray sampling and tracing (1, 3), projection transformation (2, 6), BSDF evaluation (4, 5), phase sampling (8), and complex amplitude accumulation (7, 9). Step-by-step details are presented in Sec.~\ref{sec:wavepathtracing}. \textbf{Top-right:} the wave field propagation undergoes various materials where we evaluate the BSDF of the intersected facet (point) to determine which condition, i.e., $\delta$ or non-$\delta$, to activate  before reaching the SLM, inherently encoding these complex visual cues into the hologram recording process. \textbf{Bottom-left:} we model image points as the view-dependent wave-facet to record certain angular-amplitude distribution obtained via path tracing. \textbf{Bottom-right:} wave facets are initialized with random phases sampled from Gaussian random fields to spread their energy over the desired divergence range.}} 
  \Description{Pipeline diagram with nine numbered stages for ray sampling, projection transforms, BSDF evaluation, phase sampling, and complex amplitude accumulation, with insets explaining material-dependent wave propagation, wave facets, and Gaussian random phase initialization.}
  \label{fig:method_overview}
  \vspace{-3pt}
\end{figure*}

\paragraph{Path Tracing and RS integral}
For path tracing, one can employ the rendering equation to solve for radiance under global illumination~\cite{schmidt2013path}, expressed as follows:
%\vspace{-2pt}
\begin{equation}
L_o(\mathbf{x}, \omega_o) = L_e(\mathbf{x}, \omega_o) + \int_{\Omega} f_r(\mathbf{x}, \omega_i, \omega_o) \, L_i(\mathbf{x}, \omega_i) \, \cos(\mathbf{n}, \omega_i) \, d\omega_i ,
\label{renderequation}
\end{equation}
where $L_o(\mathbf{x}, \omega_o)$ and $L_e(\mathbf{x}, \omega_o)$ denote the outgoing and emitted radiance at point $\mathbf{x}$ with the direction $\omega_o$. $L_i(\mathbf{x}, \omega_i)$ indicates the incoming radiance with the direction $\omega_i$. $f_r(\mathbf{x}, \omega_i, \omega_o)$ represents the bidirectional reflectance distribution function (BRDF).
The integral sums contributions of all possible incoming directions.
To account for both amplitude (intensity) and phase information for physical realism, one can incorporate the RS integral~\cite{buitrago2019non,shen2006fast} within the path-tracing framework to describe wavefront propagation, as:
\revise{
\begin{equation}
    u(\mathbf
    {x}) = \frac{1}{i\lambda} \iint_A u(\mathbf{x_i}) \frac{1}{r} e^{i \frac{2\pi}{\lambda} r} \cos(\mathbf{n}, \mathbf{r}) \, dS ,
\end{equation}
}
where \revise{$u(\mathbf{x})$} is the complex amplitude at point $\mathbf{x}$, $\lambda$ is the wavelength, $r$ is the distance between $\mathbf{x}$ and $\mathbf{x_i}$, and $\mathbf{r}$ denotes the vector from $\mathbf{x}$ to $\mathbf{x_i}$ yields $r = |\mathbf{r}|$. $A$ represents the illuminated aperture.

The RS integral can be interpreted as the wave-optics counterpart of Eq.~\ref{renderequation}, with Huygens’ principle enforcing isotropic radiation from each point. While this integral could be solved via a path tracing framework, it is often computationally prohibitive due to the combined costs of multi-bounce light transport and dense sampling rates inherent to wave optics.
As such, we extend the microfacet model and decompose diffuse reflection, glossy reflection, mirror reflections, and transmissions into coherent (solved by RS-integral) and incoherent components (solved by path tracing).
We then record the resulting BRDF into a complex amplitude on the surface, thereby accelerating sampling for holographic scene reconstruction.

\paragraph{Angular Spectrum of Wave Field}
For a complex amplitude $u(x, y)$, its \emph{angular spectrum} is defined as its Fourier transform $U(f_x, f_y)$, which gives the coefficients of its plane-wave decomposition. For detailed derivations, we refer readers to Supplementary Sec.~S1.1.
\revise{The key insight is that the coefficient $U(f_x, f_y)$ corresponds to the plane-wave component propagating along direction $(\theta_x , \theta_y)$, where $(\theta_x , \theta_y)=\big(\arcsin (\lambda f_x), \arcsin (\lambda f_y)\big)$. Its magnitude $|U(f_x, f_y)|$ therefore measures the strength of the wave component traveling in that direction~\cite{matsushima2003fast}, which we refer to as the \emph{angular-amplitude} in this work.}

Notably, an ideal plane wave extends infinitely in space and lacks spatial localization, while an ideal point source is perfectly localized but exhibits a spatially uniform spectrum, limiting its ability to represent anisotropic scattering.
We instead model light as \textbf{coherent plane-wave bundles} emitted from mesh facets that are locally small on the macroscopic scale yet much larger than the wavelength. Each bundle is assumed to have a finite divergence angle, allowing efficient ray-based tracing~\cite{steinberg2022towards}. This intermediate representation, lying between the extremes of an ideal point source and an ideal plane wave, can be propagated through the scene via path tracing, thereby capturing aspects of the wave--particle duality.

\paragraph{Coherent versus Incoherent Propagation}
Notably, a hologram with a smooth phase yields speckle-free images under coherent illumination, but concentrates its angular-spectrum energy near zero frequency. Consequently, after propagating through the optical system it undergoes pronounced diffraction, resulting in smaller caustics with ringing artifacts.
In real-world scenarios, light sources are generally incoherent, even if monochromatic. Objects illuminated by such sources scatter complex waves, whose initial phases fluctuate randomly over time. The human visual system integrates and averages these fluctuations over tens of milliseconds~\cite{schiffers2025holochrome,goodman2007speckle}. 
Thus, the angular-amplitude of an incoherent light beam tends to uniformly occupy the available bandwidth at the receiving plane, leading to natural defocus blur after propagation.
We leverage this context to synthesize perceptually realistic incoherent blur via time-multiplexed holograms (see also Sec.~\ref{subsection:multiframe}), an increasingly practical strategy as high-speed SLMs are gradually entering the market~\cite{choi2022time}. 

However, most aforementioned CGH algorithms are designed to generate a \emph{single} frame per inference, suggesting a substantial increase in computation time for multi-frame generation for a given 3D scene.

\section{Wave Field Rendering via Path Tracing}
\label{sec:wavepathtracing}

\revise{Figure~\ref{fig:method_overview} overviews our \method, whose rendering pipeline consists of \underline{nine} key stages. 
A ray is emitted from a point on the SLM (recording plane) in hologram space with a random initial direction (s1).
This ray is then transformed into world space (s2) and traced through the scene (s3).
Upon intersecting a facet, scattering is evaluated according to its BSDF, where different lobes are sampled probabilistically (s4).
The first non-delta BSDF sample is defined as the real image point $\mathbf{P}$ in world space. Before reaching $\mathbf{P}$, the coherent plane-wave bundle is traced with $OPL_\mathrm{world}$ accumulated. While after reaching $\mathbf{P}$, several incoherent radiance rays are traced to estimate the amplitude of the incident plane-wave bundle (s5).
Then, the virtual image point $\mathbf{P}'$ is calculated and mapped back into hologram space $\mathbf{Q}$ via a coordinate transformation (s6).
Based on the position of $\mathbf{Q}$, the $OPL_\mathrm{Holo}$ is determined (s7) and several random phase samples are selected from the pre-computed Gaussian random fields (s8) to model the incoherent scattering at $\mathbf{P}$.
Finally, the contribution of each plane-wave bundle is calculated by combining the amplitude (from s5) and phase (from s7 and s8), and then coherently accumulated on the hologram recording plane (s9).}

In the following, we introduce the wave‑optics‑based scattering model: wave facet (Sec.~\ref{subsection:wavefacet}), wave path tracing scheme for coherent plane‑wave bundles (Sec.~\ref{subsection:wavetracing}), parallel wave‑field rendering \revise{and Gaussian random fields} for multi‑frame holography (Sec.~\ref{subsection:multiframe}), and strategies for accelerated sampling (Sec.~\ref{subsection:wavesampling}). More algorithmic details are presented in Supplementary Material.
\vspace{-3pt}

\subsection{Wave Facet: Extended Microfacet Modeling}
\label{subsection:wavefacet}

The microfacet model, commonly used in physically based rendering, provides an effective description of incoherent irradiance and is formulated as a BSDF in conventional path tracing. However, for wave field rendering grounded in coherent scalar diffraction, an extended microfacet model must also incorporate phase information, not just intensity. In this work, we denote our facet model for CGH rendering as \textbf{wave facet}.

We refer to a small macroscopic surface patch (characteristic size $M \gg \lambda$) on the recording plane or object surface as a facet, with light wavelength $\lambda$.
The recording plane physically corresponds to the hologram (i.e., SLM) plane. Due to the finite feature size of the hologram, the bandwidth of a facet on the recording plane is limited: each facet can record only incident waves with angles smaller than $\arcsin \left[\lambda / (2d)\right]$ without aliasing, where $d$ is the hologram feature size.
When a coherent plane-wave bundle interacts with a facet, the scattered field may contain both coherent and incoherent components. For coherent scattering (e.g., specular reflection or ideal refraction), the wavefront remains phase-consistent in accordance with Huygens’ principle. \revise{It is therefore necessary to continue tracing the subsequent OPL to preserve correct depth cues and multi-view consistency for objects observed through specular surfaces.}
In contrast, for diffuse and glossy surfaces, photons undergo path-length variations due to stochastic scattering events. We model this incoherent behavior by applying random phase perturbations at the surface, effectively decorrelating the outgoing field. In subsequent tracing we omit further phase (OPL) accumulation---as the phase has already been randomized---and only record the ray-carried amplitude (i.e., the square root of average radiance).

\begin{figure}[tb]
  \includegraphics[width=\linewidth]{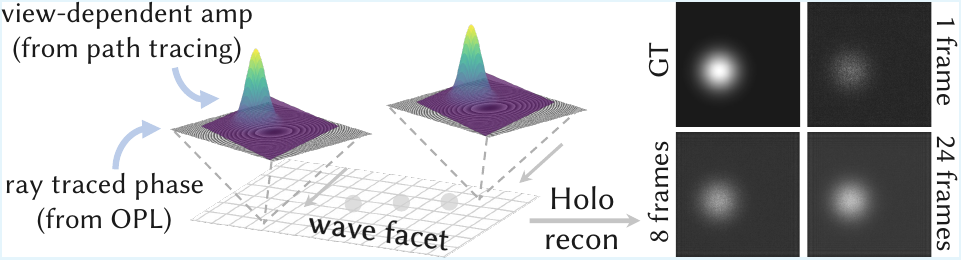}
  \vspace{-15pt}
  \caption{\textbf{Simulated angular amplitude on each wave facet}, constructed by combining the amplitude from path tracing and the phase from OPL, facilitating variable-frame-multiplexing (right).}
  \Description{Visualization of angular-amplitude distributions on representative wave facets, comparing the directional energy patterns produced by path-traced amplitudes and optical-path-length phases for multi-frame reconstruction.}
  \label{fig:ASamplitude}
  \vspace{-16pt}
\end{figure}

Matsushima et al.~\shortcite{matsushima2009extremely} show that, for both diffuse and glossy reflectance, assigning a certain random phase pattern can effectively modulate the propagation directions of the resulting wave field.
\revise{Motivated by this, we associate each rough facet in the scene (e.g., diffuse and/or glossy) with a random phase sampled from a Gaussian random field (see Sec.~\ref{subsection:multiframe} for details)}. This stochastic phase perturbation broadens the angular bandwidth of the emitted wave, yielding a statistically uniform yet random \emph{angular-amplitude}.
The degree of randomness provides an explicit control to balance angular spread against speckle noise. A subsequent path tracer then resolves the geometric component, determining how radiant energy is distributed across outgoing directions from each rough facet.
In Fig.~\ref{fig:ASamplitude} we analyze the \emph{angular-amplitude} computed by our framework, which characterizes the distribution of light's propagation directions on the wave facet.
Specifically, for light incident on a rough-like surface, the recorded \emph{angular-amplitude} closely follows the PDFs of outgoing directions for different materials, validating that the proposed wave facet model can faithfully reconstruct the material PDFs by combining the ``random-like'' phase structure and view-dependent amplitude distribution of the wave field.

\subsection{Wave Path Tracing}
\label{subsection:wavetracing}

The recording plane resides in \emph{hologram space}, where physical reconstruction is ultimately realized.
Accordingly, all diffraction accumulation for solving the RS integral must be performed in hologram space.
In contrast, path tracing is carried out in \emph{world space}, which hosts the virtual scene and is required to recover physically correct light transport.
The transformation from world space to hologram space can be achieved by $\mathbf{x}_\mathrm{holo} = \mathbf{T}\mathbf{x}_\mathrm{world}$, where $\mathbf{T}$ is defined as:
\begin{equation}
\mathbf{T} = (\mathbf{P}_\mathrm{holo}\mathbf{V}_\mathrm{holo}\mathbf{M}_\mathrm{holo})^{-1} \mathbf{P}_\mathrm{world}\mathbf{V}_\mathrm{world}\mathbf{M}_\mathrm{world},
\label{eq:transformT}
\end{equation}
where $\mathbf{M}$, $\mathbf{V}$, $\mathbf{P}$ denote model, view, and projection transformations.

Rays are randomly sampled within a limited divergence cone at the recording plane in hologram space, transformed into the world space, and then traced. 
Upon intersecting a wave facet, the path tracer, i.e., Mitsuba, samples one lobe of the material’s BSDF according to its probability. Each sampled lobe is marked as either $\delta$ or non‑$\delta$ (see top-right in Fig.~\ref{fig:method_overview}): $\delta$ lobes represent specular reflection or ideal refraction, where the outgoing direction is described by a delta function; non‑$\delta$ lobes correspond to diffuse or glossy reflection, where the outgoing direction follows a PDF determined by material properties such as metallicity and roughness.

When a ray encounters a $\delta$ lobe, we continue to propagate that path coherently by accumulating its OPL, yielding $\mathrm{OPL} = \Sigma_i L_i n_i$, where $n_i$ denotes the refractive index of the medium and $L_i$ specifies the associated segment length.
In contrast, for non-$\delta$ lobes we only compute the subsequent ray intensity, since rough surfaces induce stochastic path-length variations. Equivalently, the outgoing field can be regarded as the superposition of many coherent plane-wave bundles with random phases, making phase tracking unnecessary; therefore, we switch to incoherent path tracing to estimate the remaining intensity transport. This design improves efficiency by omitting wave sampling requirements in subsequent transport. 

Specifically, we define the point at which a ray first hits a non‑$\delta$ facet as the \emph{real image point} $P$.
The amplitude traced beyond this point is accumulated at $P$, assigned several random phases (for time-multiplexed reconstruction, see Sec.~\ref{subsection:multiframe}) to model interactions with rough surfaces, and then coherently propagated back to the recording plane.
We assume the accumulated OPL through all $\delta$‑lobes prior to $P$ to be $d_{\mathrm{OPL}}$. The \emph{virtual image point} for perception is therefore $P' = O + d_{\mathrm{OPL}}\mathbf{v}$, where $O$ and $\mathbf{v}$ denote ray origin and direction.
To compute the wave perturbation induced by $P'$ in hologram space and accumulate its contribution on the recording plane, we transform $P'$ into hologram space using Eq.~\ref{eq:transformT}, yielding $Q = \mathbf{T}P'$.
After assigning the corresponding phase and amplitude, the resulting wave is propagated to the recording plane. Summing the contributions of all sampled rays on the recording plane yields a solution to the RS integral.

\begin{figure}[tb]
  \includegraphics[width=\linewidth]{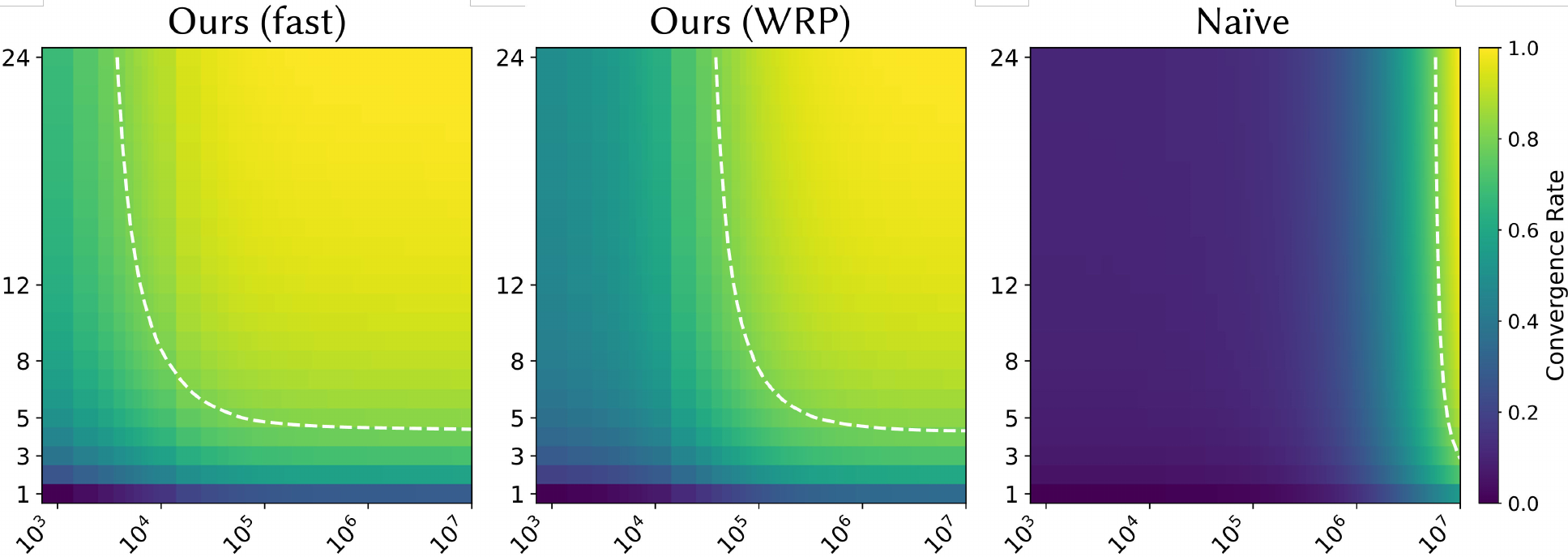}
    \vspace{-15pt}
  \caption{\textbf{Convergence behavior of wave field rendering schemes}. Our method adopts a two-stage scheme that first renders the complex-valued wave field to a wave recording plane, then uses ASM to continue propagation to the SLM (hologram) plane; we further accelerate this via texture-baking (``fast'' variant). A naïve baseline directly renders from the scene to the (typically distant) SLM plane, leading to higher computational complexity. $x$-axis: samples per pixel (SPP); $y$-axis: number of time-multiplexed frames. Convergence rate is measured by the ratio of reconstruction PSNR against that of $10^7$ spp. Note, white dashed lines indicate the 80\% convergence.}
    \Description{Convergence plots comparing direct hologram-plane rendering, the proposed two-stage wave-recording-plane scheme, and the fast texture-baked variant across samples per pixel and numbers of time-multiplexed frames, with dashed lines marking 80 percent convergence.}
    \vspace{-16pt}
  \label{fig:convergence_speed}
\end{figure}

Even after leveraging incoherent light transport to accelerate random wave propagation, the wave path tracer may still converge slowly (see Fig.~\ref{fig:convergence_speed}) due to multi-bounce evaluation for amplitude and dense RS-integral sampling for phase. To reduce amplitude variance, we adopt a texture-baking strategy~\cite{knodt2023joint} that caches view-independent ambient illumination (Supplementary Sec.~S1.4); when a ray hits the ambient-light lobe, its intensity is directly fetched from the cache. To further accelerate phase accumulation, we introduce a \emph{two-stage} strategy that substantially reduces the RS-integral sampling requirements (refer to Sec.~\ref{subsection:wavesampling}).

\begin{figure}[tb]
  \includegraphics[width=\linewidth]{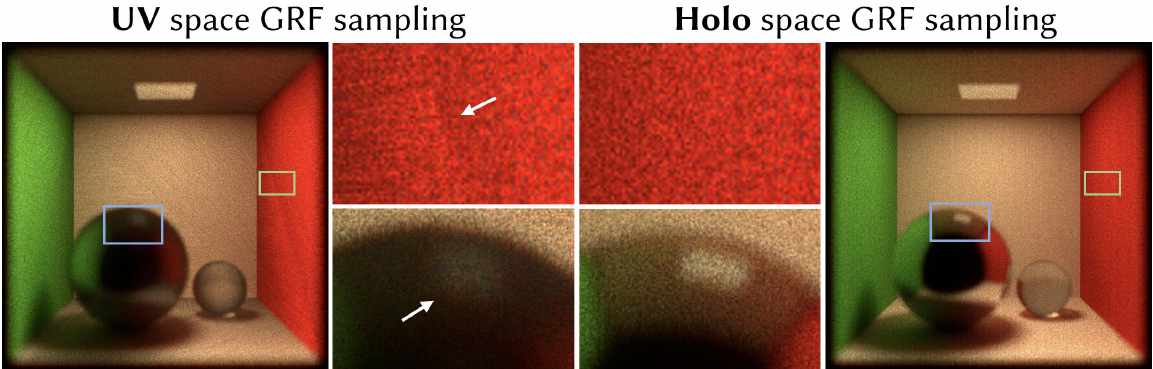}
    \vspace{-15pt}
  \caption{\textbf{Gaussian random field sampling on facet} in UV and Holo spaces.}
  \Description{Side-by-side schematic comparing random phase sampling in facet UV coordinates and hologram-space coordinates, illustrating how hologram-space sampling keeps phase bandwidth consistent across projected facets.}
  \label{fig:randomphasesampling}
    \vspace{-9pt}
\end{figure}

\subsection{Multi-frame Wave Field Rendering}
\label{subsection:multiframe}

Recent advances in digital holography with SLMs leverage multi-\allowbreak frame time-\allowbreak multiplexing to realize temporally incoherent superposition, thereby reducing coherent speckle and improving resilience to artifacts~\cite{choi2022time, chu2025artifact}. To support such a multi-\allowbreak frame holography regime, we efficiently render multiple complex wave fields of the same scene in parallel, each differing only in the random phase assigned at non-$\delta$ facets.
\revise{We note that the random phase should not arbitrarily perturb the wavefront in object-space parameterizations; instead, it should redistribute energy within the angular bandwidth physically recordable by the hologram. 
Let $\theta_{\max}=\arcsin \left[\lambda/(2d)\right]$ denote the maximum incident angle supported by the recording plane. For the 
$m$‑th frame, we therefore precompute a band-limited Gaussian random field (GRF) on the hologram plane as:
\begin{equation}
\begin{aligned}
r_m(x,y)
&=\iint_{\Omega_f} e^{i\xi_m(f_x,f_y)} e^{2\pi i(f_x x + f_y y)} \, \mathrm{d}f_x \, \mathrm{d}f_y \\
&=\mathcal{F}^{-1}\left\{\chi_{\Omega_f}(f_x,f_y)\, e^{i\xi_m(f_x,f_y)}\right\},
\end{aligned}
\end{equation}
where $\Omega_f = \left\{(f_x,f_y)\mid f_x^2+f_y^2 \leq \left(\sin \theta_{\max}/\lambda\right)^2\right\}$ indicates the admissible angular-spectrum support, $\chi_{\Omega_f}$ is the corresponding circular passband, and $\xi_m(f_x,f_y)\sim \mathcal{U}[0,2\pi]$ are independent random spectral phases. This formulation can be interpreted as a superposition of many plane-wave components with random initial phases, while confining all energy to the desired divergence cone. 
Although the Fourier-domain phases are sampled uniformly in our implementation, the inverse transform sums many independent phases, so $r_m$ is approximately Gaussian in the spatial domain by the central limit theorem; we therefore refer to it as a GRF. We then use $\varphi_m(x,y)=\arg\left(r_m(x,y)\right)$ as the random phase map of the $m$-th frame.}

\revise{Herein we consider an \emph{image point} in world space that is transformed to a point $Q(x_Q, y_Q, z_Q)$ in hologram space.  For the 
$m$‑th frame, the phase assigned to this point is sampled from the GRF at its projected location, i.e., $\phi_{Q,m}=\varphi_m(x_Q, y_Q)$. If the traced path associated with $Q$ yields amplitude $A_Q$ and optical path length $d_{\mathrm{OPL},Q}$, its contribution to the recording plane is
\begin{equation}
U_m \leftarrow U_m + A_Q \exp\left[i\left(\frac{2\pi}{\lambda}d_{\mathrm{OPL},Q} + \phi_{Q,m}\right)\right].
\end{equation}}
This yields spatially and angularly consistent random phases across the scene, as shown in Fig.~\ref{fig:randomphasesampling}. Compared with assigning phases via facet $UV$ coordinates, our strategy keeps the sampling density and effective bandwidth tied to hologram space, avoiding distortions caused by changes in facet normals or UV scaling.

Moreover, only $\phi_{Q,m}$ depends on the frame index $m$; the traced geometry, amplitudes, and OPL terms are shared by all frames. The same ray paths can therefore be reused across time-multiplexed holograms with negligible additional overhead (see Fig.~\ref{fig:speed}). 
\revise{Further discussion on path reuse is provided in Supplementary Sec.~S1.5.}

\subsection{Wave Sampling with Rays}
\label{subsection:wavesampling}
In FFT-based diffraction, the sampling requirements of the quadratic phase term depend on both propagation distance and angle~\cite{Wei2023ModelingOffaxisDiffraction}, and must be performed on a specific plane.
If the pre-defined spatial or angular sampling intervals are coarser than required, numerical diffraction produces artifacts.
In contrast, our path tracing-based approach avoids predefined sampling grids by employing Monte Carlo ray sampling, equivalent to randomly sampling the sub-hologram of each object point. This strategy suppresses artifacts of undersampling and naturally supports heuristic, spatially varying non‑uniform sampling schemes.

Converting rays to wave fields on the hologram plane resembles an integral-based diffraction operation with computational complexity \( \mathcal{O}\left(N_{\text{p},0} N_{\text{ray},0}\right) \), where \(N_{\text{p},0}\) and \(N_{\text{ray},0}\) denote the pixel samples and ray samples per pixel (SPP) on the hologram plane, respectively.
Here, SPP corresponds to the number of samples in the angular domain and scales quadratically with the hologram area, which itself grows quadratically with propagation distance. 
Consequently, the overall complexity of this Monte Carlo integral increases approximately with the fourth power of propagation distance.
To obtain a high-fidelity hologram, it is essential to satisfy these sampling requirements during path tracing.
In our setting, the hologram plane is placed \qty{85}{\mm} away from the farthest scene point.
To reduce the computational cost, we adopt a \emph{two-stage} strategy inspired by the wave recording plane (WRP) technique~\cite{Shimobaba2009SimpleFastCalculation}.
Specifically, we first convert traced rays into wave fields on a WRP placed \qty{5}{\mm} away from the farthest scene point, then propagate this field from WRP to the hologram plane using ASM.
The computational complexity of this procedure is \( \mathcal{O}\left(N_{\text{p}} N_{\text{ray}} + 2 N_{\text{p}}\log N_{\text{p}}\right) \), where \(N_{\text{p}}\) and \(N_{\text{ray}}\) denote the counterparts on WRP.
The factor 2 accounts for double zero-padding in FFT-based ASM to avoid aliasing.
In practice, the ASM overhead is negligible because both \(N_{\text{ray}}\) and \(N_{\text{p}}\) remain much smaller than their counterparts \(N_{\text{ray},0}\) and \(N_{\text{p},0}\) on the distant hologram plane.
Therefore, as illustrated in Fig.~\ref{fig:convergence_speed}, compared to direct conversion on the hologram plane, this WRP-based scheme substantially reduces the required samples. 
which is particularly beneficial when enforcing the multi-frame wave field rendering in practical CGH. 
\vspace{-2pt}

\begin{figure}[tb]
  \includegraphics[width=0.85\linewidth]{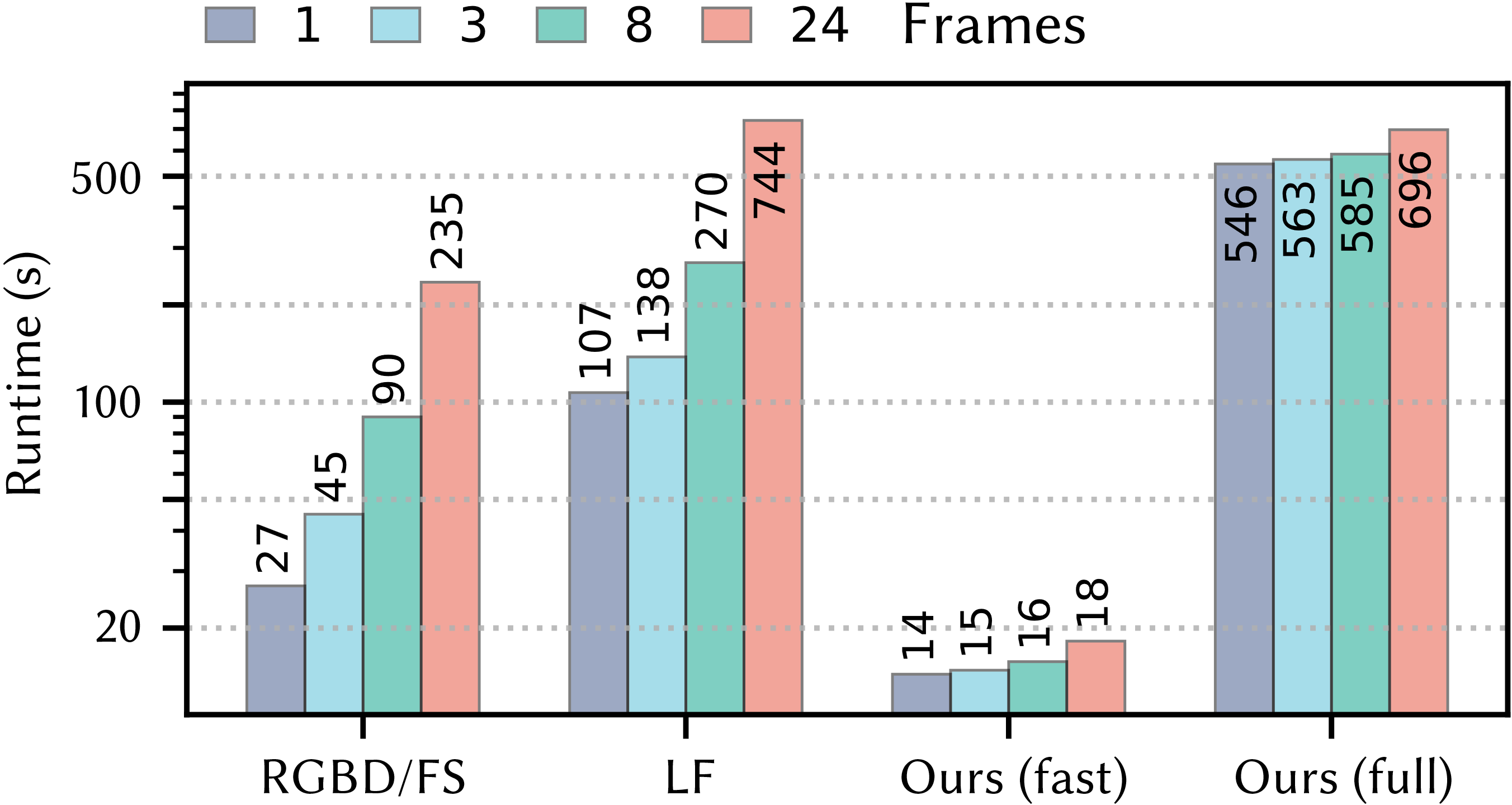}
    \vspace{-6pt}
  \caption{\textbf{Runtime ablation} on representative CGH frameworks. Colors indicate varying-frame-multiplexing: 1, 3, 8, 24 frames. The proposed methods generate multiple frames with only marginal computational overhead.}
  \Description{Runtime comparison chart for representative CGH frameworks under 1, 3, 8, and 24 time-multiplexed frames, showing that the proposed methods add little overhead as the frame count increases.}
  \label{fig:speed}
    \vspace{-21pt}
\end{figure}

\section{Experiments and Results}

\vspace{-1pt}
\subsection{Scene and System Configurations}

We use readily available 3D scenes~\cite{benedikt16resources} from Mitsuba, which we further edit in Blender to show complex rendering effects. Their visualization is presented in Supplementary Sec.~S4.
The wavelengths for simulations are set to \SI{640.0}{\nano\meter} (red), \SI{516.5}{\nano\meter} (green), and \SI{455.4}{\nano\meter} (blue), with a pixel pitch of 8~$\mu$m and a propagation distance of 80~mm (corresponding to virtual infinity). \revise{Detailed setup specifications are provided in Supplementary Sec.~S5. Baseline methods for assessment are described in Supplementary Sec.~S2.}

\subsection{Rendering Results}
\begin{figure*}
  \includegraphics[width=0.96\textwidth]{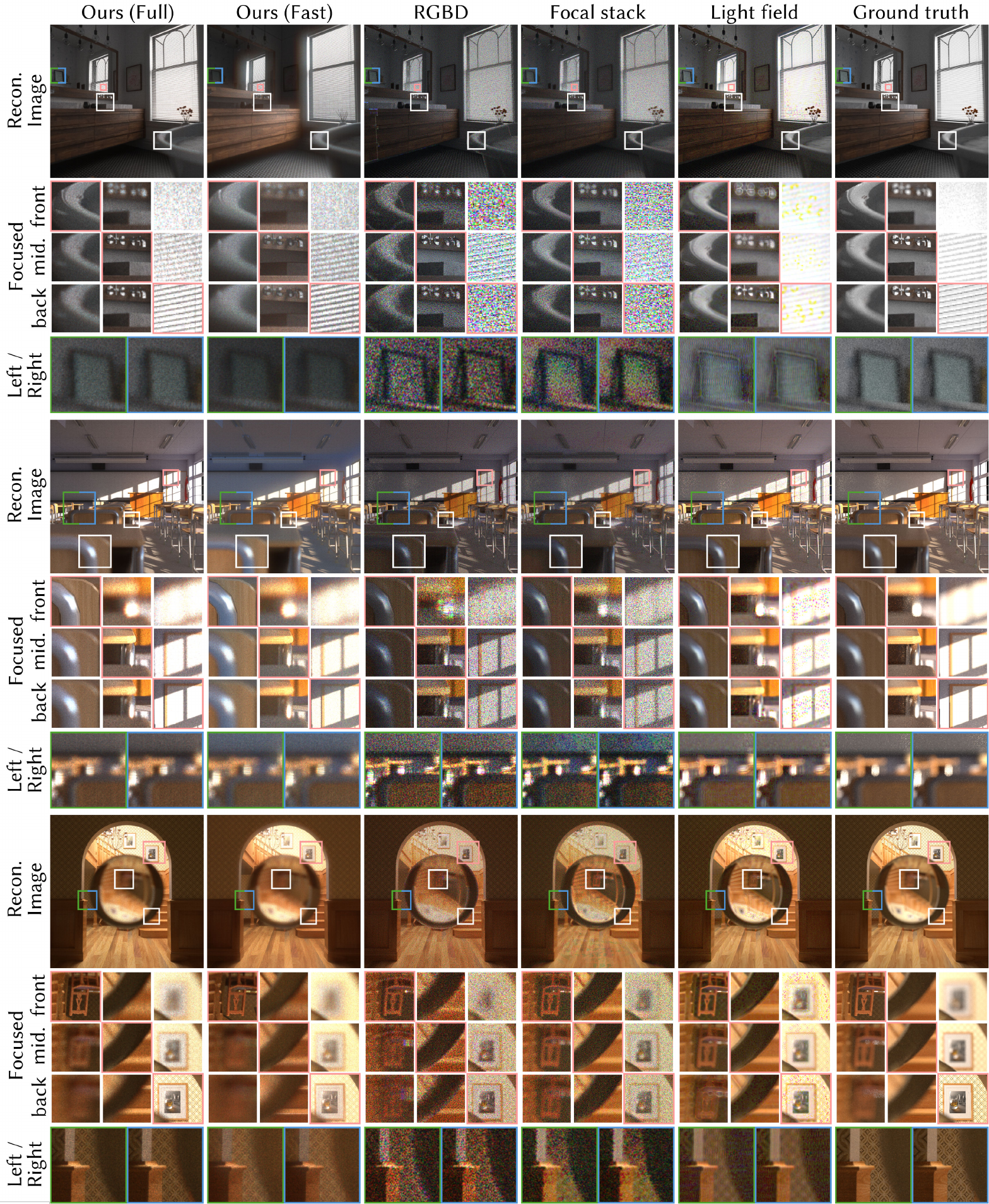}
  \vspace{-6pt}
  \caption{\textbf{Rendering results of different CGH frameworks}. \emph{Fast} denotes our rendering using texture-baking for computation acceleration.}
  \Description{Grid of simulated reconstruction results for several CGH frameworks across multiple scenes, with close-up regions comparing depth-dependent focus, defocus, and visual fidelity; the fast variant denotes the texture-baked version of our method.}
  \label{fig:results_simulation_1}
\end{figure*}

\begin{figure*}[htb]
  \includegraphics[width=1\textwidth]{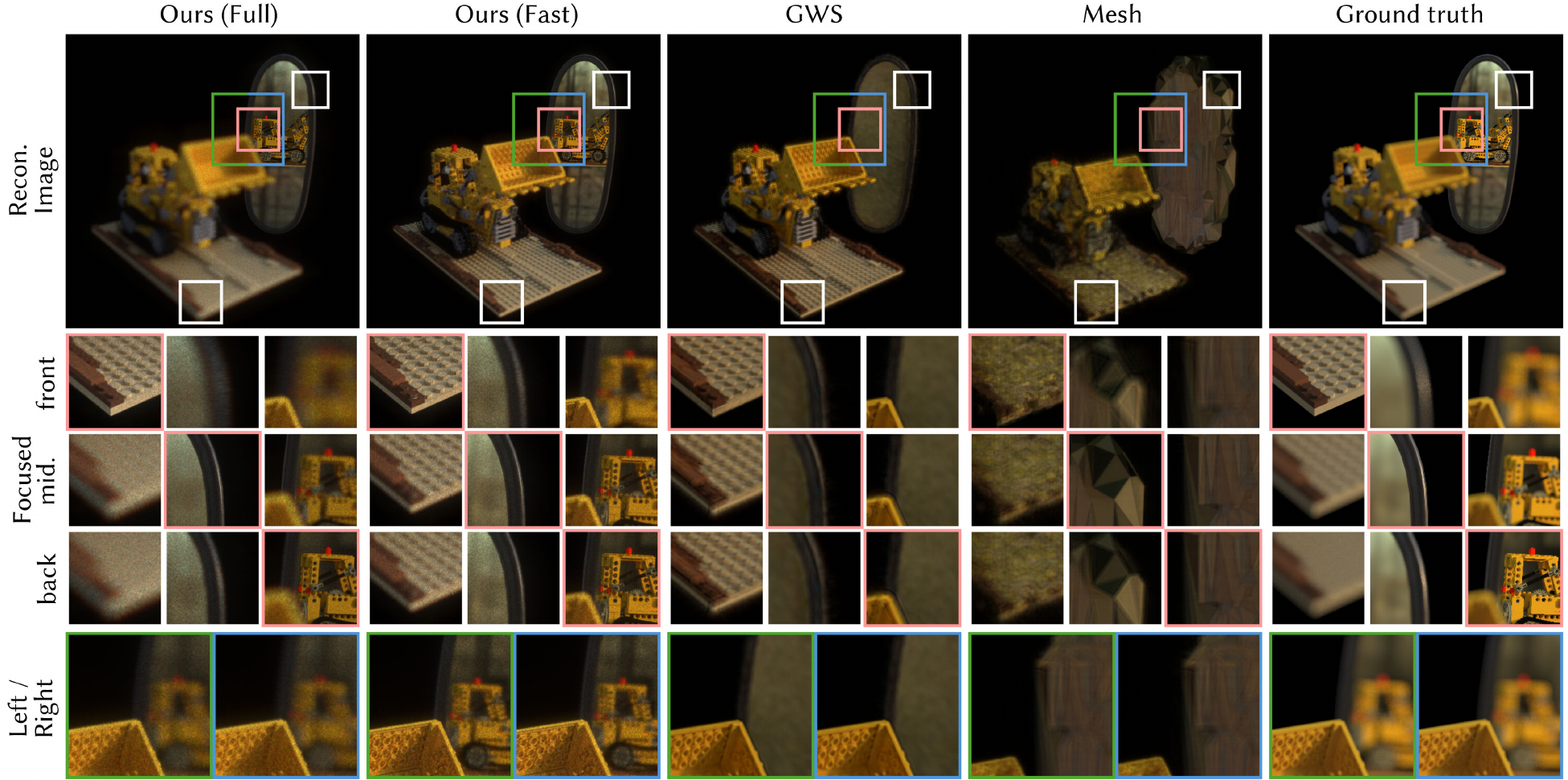}
    \vspace{-18pt}
  \caption{\textbf{Rendering results} of our two variants, Gaussian wave splatting, and mesh-based. We refer readers to ``imagery'' differences on the \emph{mirror}.}
  \Description{Comparison of simulated reconstructions from the full and fast variants of HoloPathTracer, Gaussian wave splatting, and mesh-based CGH, emphasizing differences in mirror imagery and depth-dependent appearance.}
  \label{fig:results_simulation_2}
    % \vspace{-2pt}
\end{figure*}

Figure~\ref{fig:results_simulation_1} compares rendering results of different CGH frameworks, including two HoloPathTracer variants, RGBD, focal stack (FS), light field (LF), and ground truth from Mitsuba3 (GT). ``Fast'' denotes our renderings employing a texture-baking strategy for accelerated computation. Similarly, Fig.~\ref{fig:results_simulation_2} compares rendering results of Ours, Gaussian wave splatting (GWS), and mesh-based (Mesh). For each set, we present close-up regions with the viewpoint focused at front, middle, and back, as well as the left/right view-dependent effects. 

Across evaluated scenes, RGBD exhibits noise-corrupted defocus regions (e.g., when focused at front) due to depth-layer discontinuities in supervision.
FS improves the defocus consistency but softens details in focused regions (see red boxes for middle-focus).
LF provides the highest visual quality among image/layer-based baselines because of dense angular supervision, yet noticeable artifacts remain (e.g., third-column close-ups), likely due to overfitting to predefined viewpoints and pupil sizes.
Both GWS and Mesh struggle to reproduce accurate mirror-like visual cues (see pink boxes where Lego truck imagery is missing in the mirror).
Our method closely matches GT under arbitrary viewpoints, refocus distances, and pupil sizes, yielding smooth focus transitions, and accurately reproducing lens/mirror-formed imagery.
``Fast'' variant preserves the same focus cues while being faster than baselines, with only minor detail/color drifts mainly due to baking approximations. \revise{Please refer to the Supplementary Material for results on additional scenes, and to the Supplementary Video, where the view-dependent effects and focus transitions are more clearly visible.}

%%%%%%%%%%%%%%%%%%%%%%%%%%%%%%%%%%%%%%%%%%%
\vspace{-2pt}
\subsection{Experimental Results}

\paragraph{Phase-only Hologram Encoding.}
We adopt the established SGD-based optimization following recent literature~\cite{choi2021neural}. Importantly, to better align with the scope of efficiently and accurately encoding both amplitude and phase, we supervise the optimization directly in the complex-valued domain, rather than supervising only the amplitude as in the prior work.
\revise{In such a way, one would only need to supervise on one single reconstruction plane without losing both depth and view continuity, significantly reducing optimization time.} Further details of the phase-only optimization are described in Supplementary Sec.~S3.2.

\paragraph{Physical Configuration.}
Our holographic display prototype closely follows the simulation, as shown in Supplementary Fig.~S4. 
Synthesized holograms are quantized and displayed on a phase-only SLM,
illuminated by a fiber-coupled RGB laser. The beam is directed through a neutral density filter, a collimating lens, a linear polarizer, and a beam splitter before incident onto the SLM. The modulated wavefronts transmit through an eyepiece and a lens before being recorded by a sensor.
We configure an iris mounted on translation mechanics in between the eyepiece and the camera lens to mimic the pupil-filtering.
For multi-depth holographic reconstruction, we propagate to 5 depth planes, uniformly spaced at $1.25$~mm intervals within 80--85~mm.
The camera focus can be adjusted using an Arduino micro-controller to support re-focusing on multiple pre-calibrated planes. At this proof-of-concept stage, color and time-multiplexing results are obtained via post-image processing (e.g., 24 frames).
Detailed specifications of devices/parts are presented in Supplementary Sec.~S5.

\begin{figure*}[h]
\centering
    \includegraphics[width=1\textwidth]{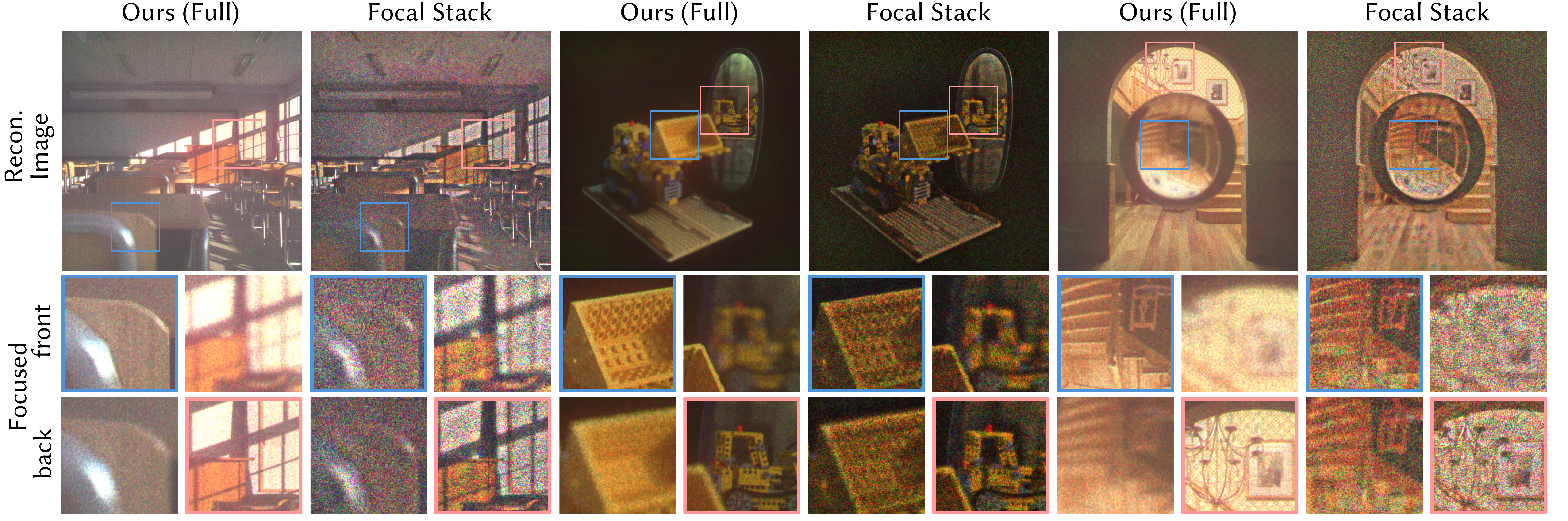}
    \vspace{-18pt}
  \caption{\textbf{Acquired display results} comparing Ours with Focal Stack, with front- and back-focused close-ups. Our method produces sharper in-focus details and more natural defocus. The observed contrast loss primarily stems from the higher sensitivity of our random-phase holograms to SLM crosstalk compared with smoother, single-view-supervised image-based holograms, while better supporting multi-view reconstruction.}
  \Description{Photographically captured focal-stack reconstructions from the holographic display prototype, comparing HoloPathTracer and a focal-stack baseline with front-focused and back-focused zoomed regions across several scenes.}
  \label{fig:results_experiment}
  \vspace{-6pt}
\end{figure*}

\paragraph{Acquired Display Results.}
We present captured 3D focal stacks of holograms generated and displayed on our experimental prototype in Fig.~\ref{fig:results_experiment}, comparing ours with the representative focal stack (FS) baseline.
For each set, we present close-up regions highlighting depth-dependent effects. We observe across three scenes that ours produced sharper in-focus details and more natural defocus effects.

\revise{Noteworthy, although the displayed image contrast may appear low, it remains comparable to that reported in recent CGH systems~\cite{chakravarthula2022hogelfree,kuo2023multisource,chu2025artifact} built upon similar light-engine setups without extensive filters. In addition, random-phase holograms are known to be more sensitive to SLM inter-pixel crosstalk, which can further reduce image contrast. Possible remedies include learned camera-in-the-loop calibration~\cite{choi2021neural}, crosstalk-aware optimization~\cite{markley2023simultaneous}, and higher-fidelity SLM hardware. Improving display contrast is therefore largely orthogonal to the main scope of this work, which is to establish a fast wave path tracing framework that faithfully reproduces 3D visual cues.}

\section{Discussion and Conclusion}

We develop a path tracing-based wave optics rendering paradigm for CGH that encodes physically accurate 3D scene attributes into holograms while maintaining reasonably fast computation.
In particular, our method delivers physically accurate and spatially continuous depth cues, even in the presence of curved reflective and/or refractive surfaces. We further achieve substantial computation speedups via coherency decomposition, ambient-light caching, and a two-stage sampling strategy. We envision this novel wave field rendering framework paving the way towards efficient, multi-frame, random hologram generation in large-scale, complex 3D scenes.
We validate our HoloPathTracer framework through simulations and experiments, demonstrating photorealistic reconstruction quality comparable to state-of-the-art scene rendering pipelines (e.g., the Mitsuba renderer) on off-the-shelf scene datasets.
%\vspace{-2pt}

\noindent\paragraph{Limitations and Future Work}
While our wave field rendering execution has been demonstrated to efficiently support
multi-frame hologram generation with only marginal computational overhead, its runtime is presently bounded by the inherent sampling cost of wave propagation, and can be further accelerated by training advanced network models to represent/infer the propagation (rendering)~\cite{zeng2025renderformer,qu202583}.
We also observe minor visual discrepancies in brightness (dynamic range) and color details between our reconstruction and ground-truth, especially for the ``fast'' variant tailored for higher computational efficiency.

\revise{For scenes illuminated by incoherent light sources, the time-multiplexed random phases reconstruction adopted in this work is an unbiased estimator of the target light-field intensity distribution in expectation. The remaining discrepancies are therefore mainly tied to modeling approximations in the implementations. In particular, the binary $\delta$/non-$\delta$ treatment can introduce errors for low-roughness glossy materials that still preserve limited imaging ability, and using a single wave recording plane may cause color drift for objects placed very close to it due to the limited spatial resolution for recording view dependence. Roughness-aware coherence modeling and multiple recording planes are promising directions to mitigate these effects, as further discussed in Supplementary Sec.~S1.3.}

In addition, our experimental holographic display quality is bound\-ed by the restricted \'etendue of the SLM and residual degradations from imperfect optics. 
The \'etendue bottleneck is a long-standing challenge in CGH, 
and emerging photonics-based SLMs with much smaller pixel pitch~\cite{kaczorowski2024sub} promise to lift this limitation. 
A camera-calibrated, learned propagation model, similar to those explored in prior work~\cite{choi2021neural}, could correct for most residual degradations along the optics path. We regard such calibration as a valuable future add-on.


\vspace{-3pt}
\begin{acks}
We thank Jinfan Lu and Li Liao for fruitful discussions. This work was partially supported by the National Natural Science Foundation of China (62322217), the Innovation and Technology Fund of Hong Kong (MHP/313/24), and the Research Grants Council of Hong Kong (GRF 17208023).
\end{acks}

%%
% \newpage
\bibliographystyle{ACM-Reference-Format}
\bibliography{ref}

% \appendix

\end{document}